\documentclass[pmlr]{jmlr}% new name PMLR (Proceedings of Machine Learning Research)

\usepackage{longtable}% for long tables
\usepackage{booktabs}
 \usepackage{siunitx}

\theorembodyfont{\upshape}
\theoremheaderfont{\scshape}
\theorempostheader{:}
\theoremsep{\newline}

\jmlrvolume{(PREPRINT Submitted to PMLR)}
\jmlryear{}
\jmlrworkshop{}

\title[NeurIPS2021 BEETL Competition (PREPRINT)]{2021 BEETL Competition: Advancing Transfer Learning for Subject Independence \& Heterogenous EEG Data Sets}
\author{
\Name{Xiaoxi Wei}\textsuperscript{a} \Email{xiaoxi.wei18@imperial.ac.uk}\\ 
\Name{A. Aldo Faisal\textsuperscript{a,b}} \Email{aldo.faisal@imperial.ac.uk}\\
\Name{Moritz Grosse-Wentrup\textsuperscript{c}} \Email{moritz.grosse-wentrup@univie.ac.at}\\
\Name{Alexandre Gramfort\textsuperscript{d}} \Email{alexandre.gramfort@inria.fr}\\
\Name{Sylvain Chevallier\textsuperscript{e}} \Email{sylvain.chevallier@uvsq.fr }\\
\Name{Vinay Jayaram\textsuperscript{f}} \Email{vinayjayaram@fb.com}\\
\Name{Camille Jeunet\textsuperscript{g}} \Email{camille.jeunet@u-bordeaux.fr}\\
\addr \textsuperscript{a}Brain {\&} Behaviour Lab, Imperial College London, United Kingdom\\
\textsuperscript{b}Institute of Artificial \& Human Intelligence, University of Bayreuth, Germany\\
\textsuperscript{c}Faculty of Computer Science, CogSciHub, Data Science@Univie, University of Vienna, Austria\\
\textsuperscript{d}Universite Paris-Saclay, Inria, CEA, Palaiseau, France\\
\textsuperscript{e}LISV, UVSQ, Université Paris-Saclay, France\\
\textsuperscript{f}Reality Labs, USA\\
\textsuperscript{g}University of Bordeaux, France\\
\AND
\Name{Stylianos Bakas}\textsuperscript{1}$^,$\textsuperscript{2}$^,$\textsuperscript{3} \Email{stelios@cogitat.io}\\
\Name{Siegfried Ludwig}\textsuperscript{1}$^,$\textsuperscript{2} \Email{siegfried@cogitat.io}\\
\Name{Konstantinos Barmpas}\textsuperscript{1}$^,$\textsuperscript{2} \Email{ntinos@cogitat.io}\\
\Name{Mehdi Bahri}\textsuperscript{1}$^,$\textsuperscript{2} \Email{mehdi@cogitat.io}\\
\Name{Yannis Panagakis}\textsuperscript{1}$^,$\textsuperscript{2}$^,$\textsuperscript{4} \Email{yannis@cogitat.io}\\
\Name{Nikolaos Laskaris}\textsuperscript{1}$^,$\textsuperscript{2}$^,$\textsuperscript{3} \Email{nikos@cogitat.io}\\
\Name{Dimitrios A. Adamos}\textsuperscript{1}$^,$\textsuperscript{2}$^,$\textsuperscript{3} \Email{dimitrios@cogitat.io}\\
\Name{Stefanos Zafeiriou}\textsuperscript{1}$^,$\textsuperscript{2} \Email{stefanos@cogitat.io}\\
\Name{William C. Duong}\textsuperscript{5}$^,$\textsuperscript{6}
\Email{wduong@dcscorp.com}\\
\Name{Stephen M. Gordon}\textsuperscript{5}$^,$\textsuperscript{6}
\Email{sgordon@dcscorp.com}\\
\Name{Vernon J. Lawhern}\textsuperscript{6} \Email{vernon.j.lawhern.civ@army.mil}\\
\Name{Maciej Śliwowski}\textsuperscript{7}$^,$\textsuperscript{8}$^,$\textsuperscript{9} \Email{maciej.sliwowski@opium.sh}\\
\Name{Vincent Rouanne}\textsuperscript{7} \Email{vincent.rouanne@gmail.com}\\
\Name{Piotr Tempczyk}\textsuperscript{9}$^,$\textsuperscript{10} \Email{piotr.tempczyk@opium.sh}\\
\addr \textsuperscript{1}{Cogitat Ltd., United Kingdom}\\
\textsuperscript{2}{Intelligent Behaviour Understanding Group, Imperial College London, United Kingdom}\\
\textsuperscript{3}{Aristotle University of Thessaloniki, Greece}\\
\textsuperscript{4}{National and Kapodistrian University of Athens, Greece}\\
\textsuperscript{5}{DCS Corporation, Alexandria, VA, USA}\\
\textsuperscript{6}{Human Research and Engineering Directorate, DEVCOM Army Research Laboratory, Aberdeen Proving Ground, MD, USA}\\
\textsuperscript{7}{Univ. Grenoble Alpes, CEA, LETI, Clinatec, F-38000 Grenoble, France}\\
\textsuperscript{8}{Université Paris-Saclay, CEA, List, F-91120, Palaiseau, France}\\
\textsuperscript{9}{Polish National Institute for Machine Learning (OPIUM), Warsaw, Poland}\\
\textsuperscript{10}{deeptale.ai, Poland}\\
}
\editor{Editor's name}
\begin{document}
\maketitle
% \newpage
\begin{abstract}
Transfer learning and meta-learning offer some of the most promising avenues to unlock the scalability of healthcare and consumer technologies driven by biosignal data. This is because current methods cannot generalise well across human subjects' data and handle learning from different heterogeneously collected data sets, thus limiting the scale of training data. On the other side, developments in transfer learning would benefit significantly from a real-world benchmark with immediate practical application. Therefore, we pick electroencephalography (EEG) as an exemplar for what makes biosignal machine learning hard. We design two transfer learning challenges around diagnostics and Brain-Computer-Interfacing (BCI), that have to be solved in the face of low signal-to-noise ratios, major variability among subjects, differences in the data recording sessions and techniques, and even between the specific BCI tasks recorded in the dataset.
Task 1 is centred on the field of medical diagnostics, addressing automatic sleep stage annotation across subjects. Task 2 is centred on Brain-Computer Interfacing (BCI), addressing motor imagery decoding across both subjects \textit{and} data sets. The BEETL competition with its over 30 competing teams and its 3 winning entries brought attention to the potential of deep transfer learning and combinations of set theory and conventional machine learning techniques to overcome the challenges. The results set a new state-of-the-art for the real-world BEETL benchmark.
\end{abstract}


% \newpage
\section{Introduction}
\label{sec:intro}

The maturing of machine learning methods and their progressive deployments into the real-world brought to the forefront the need for combining similar data sets. Transfer learning has become a promising strategy to align different distributions \cite{pan2009survey}. Transfer learning encompasses algorithms to transfer the representations and knowledge from source domains to a target domain. In the machine learning field, fine-tuning is often used to transfer model representations between tasks or data sets \cite{11xx}. Another strategy is model splitting \cite{12xx}, which uses multiple sets of parameters for different data sets. Deep domain adaptation \cite{13xx} is a technique which directly projects features of data sets into a common space with neural networks.

We pick electroencephalography (EEG), which is broadly considered one of the most promising ways to non-invasively read out of the human brain for diagnostic and human interfacing purposes. EEG reflects the features that make biosignal understanding a hard problem, as it consists of multi-dimensional time-series data that suffers from signal non-stationarities and poor signal-to-noise ratio, variability between users and sessions, different channel numbers and locations, as well as differences in task definitions between data sets. While EEG hardware has steadily evolved, trustworthy and data-efficient decoding methods are still missing. Currently, there are a few reviews \cite{2,3,6} on different EEG transfer learning algorithms; and some recent inter-subject studies based on deep transfer learning approaches in Brain-Computer Interfaces (BCI) (e.g. \cite{wei2021inter,li2021meta}) and in sleep studies (e.g. \cite{chambon2018deep, andreotti2018multichannel}. However, the field still lacks a systematic benchmark since these algorithms are tested on different data sets or with different pre-processing and setups in their studies. More importantly, very limited work focuses on cross-dataset transfer learning, which limits the use of big data in EEG decoding.

Two challenging tasks are designed to stimulate such algorithmic innovation. Task 1 is centred on transfer learning in the field of medical diagnostics, addressing automatic sleep stage annotation. The challenge lies in transferring from a large control population data set to clinically relevant cohorts with very little training data (transfer across subjects). Task 2 is centred on transfer learning for BCI, addressing motor imagery decoding. The challenge lies in transferring from multiple data sets, which use different EEG setups comprising hundreds of users, to a set of new users that need to be up and running with only minutes worth of calibration data (transfer across subjects \textit{and} data sets). Currently, most studies focus on transfer learning within a single data set. The Benchmarks for EEG Transfer Learning (BEETL) competition provides a referencing platform for transfer learning strategies to two of the most common EEG applications and a guide on combining and utilising data sets from different sources, which could promote the EEG field towards the use of big data.

\section{Task Description}
\label{sec:task}

Task 1 is in the field of medical diagnostics and specifically has the goal of automatic sleep stage annotation from sleep EEG data. We provide a data set with adult users (40 users, age 22-65) with 6 label categories for model training, based on which sleep stage annotation has to be transferred to two different age groups (65-80 and 80+) for each of which 5 subjects worth of data are provided. Task 1 is an essential use case for the development of ready-to-use medical diagnostics developed on a standard, large user base that has to be then transferred to many different clinically relevant subpopulations, for which respectively only a few subjects are worth of data can be collected. Beyond requiring subject-independence, the transfer has to work on different user groups (elderly and very elderly subjects) with well documented systematic EEG differences during sleep \cite{landolt1996effect,
boselli1998effect,landolt2001age,purdon2015ageing}.

Task 2 is a 3-way motor imagery classification challenge (left-hand, right-hand motor imagery and 'reject') that gets at the heart of the problem of current BCI systems: motor imagery data is exhausting for subjects to record, and historically has been difficult to use in a cross-subject and cross-dataset manner. Currently, there is limited work on cross-dataset transfer learning, and existing methods lack a systematic experimental comparison in the literature. This task provides a platform to compare the performance of current transfer learning algorithms across both subjects and data sets. In the past we organized several motor imagery data sets for BCI challenges in the MOABB (Mother of All BCI Benchmarks, \url{https://github.com/NeuroTechX/moabb}) database to test the performance of algorithms in terms of their generalisation performance on new data sets. Three source data sets are provided as training data. The algorithms are evaluated on new data sets with different setups, including differences in electrode channels, task definitions, and subjects. The test set contains an unpublished data set that is collected for this purpose and will be added to the MOABB database post-competition. Demonstration figures and more details of tasks can be found on the BEETL website (\url{https://beetl.ai/challenge}).

\section{Data}
\label{sec:Data}

\textbf{Sleep Task data set.} For Task 1, the sleep stage decoding task, the Physionet sleep data set \cite{10,11} is one of the ideal data sets. The Sleep-EDF is a public database ( \url{https://physionet.org/content/sleep-edfx/1.0.0/}) that contains 197 whole-night sleep recordings with event markers annotated by experts. Sleep patterns consist of sleep stages W, R, 1, 2, 3, 4, M (Movement time) and '?' (not scored). This data set has a clustered distribution of participants of different ages. The number of subjects is large enough for transfer learning algorithms to learn the diversity of distributions. We selected and randomized the subjects and trials in the competition to avoid cheating. Processed data could be found in the competition start kits.

\noindent \textbf{Motor Imagery Task data set.} For Task 2, the motor imagery decoding task, we selected three public data sets (Cho2017 \cite{18}, BNCI2014 \cite{13} and PhysionetMI \cite{11,schalk2004bci2000}) from the MOABB database as sources (\url{http://moabb.neurotechx.com/docs/datasets.html}). MOABB is a framework for evaluating BCI classification algorithms on publicly available data sets. We have collected a data set in an online racing game format in Cybathlon2020IC \cite{wei2021inter} for testing purposes. Some offline subjects from the Weibo2014 data set \cite{weibo} and some online collected subjects from the Cybathlon2020IC data set are used as test samples. Data set information can be found in the table \ref{MIdataset}. Detailed description of the data could be found on the MOABB and BEETL websites. (\url{https://beetl.ai/data}).

\begin{table}[htbp]
\centering
\caption{MI data sets in BEETL}
\begin{tabular}{llll}
\hline
MI Data set             & Subjects & Channels & Tasks                                \\ \hline
Cho2017               & 52       & 64       & Left/Right hand                 \\
BNCI2014               & 9        & 22       & Left/Right hand/Feet/Tongue     \\
PhysionetMI            & 109      & 64       & Left/Right hand/Feet/Both hands/Rest \\ 
Weibo2014                & 10       & 60       & Left/Right hand/Feet/Rest       \\
Cybathlon2020IC & 5        & 63       & Left/Right hand/Feet/Rest \\ \hline
\end{tabular}
\label{MIdataset}
\end{table}

\noindent\textbf{Training, validation and test set.} For the Physionet sleep data set, we provide 80 sessions from 40 subjects (aged from 25-64) with full labels as source data and 5 subjects aged from 65 to 79 as examples of this age group; the performance of the algorithm is tested on more subjects aged from 65 to 79. Similarly, we provide 5 subjects aged from 80 to 95 with labels, while accuracies are reported on other subjects aged from 80 to 95. For the Physionet MI, Cho2017 and BNCI data sets, we provide full data sets with labels as sources. In both the Weibo2014 and Cybathlon2020IC data set, we provide some data with labels per test subject. During the competition, 32 channels around the motor cortex are selected from the Weibo2014 data set. The data set name was not provided during competition to avoid cheating. For the validation data (phase 1, leaderboard phase in the competition), two subjects from the Cybathlon2020IC data set and subjects 3, 4, and 5 of the Weibo2014 data set are used. For the test data (phase 2, final ranking phase in the competition), three subjects from the Cybathlon2020IC and subjects 7 and 9 from Weibo2014 are used as testing samples. Cybathlon2020IC will be added to MOABB post-competition.

\noindent\textbf{Metrics.} As both tasks in the BEETL challenge are classification problems, classification accuracy on the test data is the standard for ranking different solutions. To account for class imbalances in the data, balanced accuracy is computed by giving higher weight to classes with less samples \cite{brodersen2010balanced}. The final competition score is the sum of the respective scores on both tasks.
% As both tasks in the BEETL challenge are classification problems, classification accuracy on the test data is the standard for ranking and we give higher weights to classes with fewer samples. The weight of each predicted sample in class i is $ {Weight}_i = \\ {(100/Classes)}/  {TrueLabelSize_i} $, where Classes is 6 in Task1 and 3 in Task 2. In the final team ranking we combine the score across the two tasks as follows for each participant $i$:
% % \aldo{make this equation online, it wasts space and the math is trivial}
% $   \mbox{Final Score}_i = \mbox{Score Task 1} + 
%     \mbox{Score Task 2}
% $.\\

% \input{4Metrics.tex}
\section{Competition Results and Solutions}
\label{sec:result}

In task 1, winning teams used a single neural network for all subjects during training and testing based on DeepSleep \cite{chambon2018deep} or EEGInception \cite{santamaria2020eeg}. In task 2, the variability across different data sets, tasks, channel locations, and sampling rates make transfer more challenging (table \ref{MIdataset}). All three teams selected the common channels and aligned the sampling rate of different data sets. Team Cogitat used the Deep Sets \cite{zaheer2017deep} framework with EEGInception to align latent distributions across different data sets and subjects. Team Wduong used Label Alignment \cite{he2020latransfer} and Euclidean Alignment \cite{he2019transfer} to narrow the gap among tasks and data sets. After the alignment, team Wduong used an EEGNet \cite{lawhern2018eegnet} backbone combined with a multi-task learning setup \cite{ruder2017overview} and the Maximum Classifier Discrepancy (MCD) \cite{mcd} method to perform classification. Team ms01 performed deep transfer learning with an architecture based on Riemannian geometry approaches \cite{spdnet}. The winners and respective decoding scores on the two tasks can be found in Table \ref{ranking} (more details available at \url{https://beetl.ai/prizes}).\\

\begin{table}[htbp]
\centering
\caption{Method Accuracies of Top 5 Teams}
\begin{tabular}{llll}
\hline
{\color[HTML]{000000} }        & {\color[HTML]{000000} Task 1} & {\color[HTML]{000000} Task 2} & {\color[HTML]{000000} Overall} \\ \hline
{\color[HTML]{000000} Cogitat} & {\color[HTML]{000000} 65.55} & {\color[HTML]{000000} 76.33} & {\color[HTML]{000000} 141.88}  \\
{\color[HTML]{000000} wduong}  & {\color[HTML]{000000} 68.66} & {\color[HTML]{000000} 71.33} & {\color[HTML]{000000} 139.99}  \\
{\color[HTML]{000000} ms01}    & {\color[HTML]{000000} 65.57} & {\color[HTML]{000000} 59.87} & {\color[HTML]{000000} 125.44}  \\
{\color[HTML]{000000} nik-sm}  & {\color[HTML]{000000} 69.23} & {\color[HTML]{000000} 54.47} & {\color[HTML]{000000} 123.7}   \\
{\color[HTML]{000000} michaln} & {\color[HTML]{000000} 66.78} & {\color[HTML]{000000} 56.47} & {\color[HTML]{000000} 123.25}  \\ \hline
\end{tabular}
\label{ranking}
\end{table}

\noindent\textbf{Baseline and Start kits}
A naive baseline without transfer learning was evaluated with the Shallow ConvNet from \cite{16}. In task 1, training data of the baseline method contains the source subjects and the example subjects from the test population. In task 2, evaluation was done on each subject with their own example data to avoid the negative transfer problem reported in \cite{wei2021inter}. 17 common channels of all data sets are used in the baseline of task 2. The baseline score in task 1 and task 2 is 57.6\% and 49.9\% respectively. Start kits for loading data, training models and generating labels are provided in Python (\url{https://beetl.ai/code}).

\subsection{First Place Solution: Latent Subject Alignment}

Team \verb|Cogitat|: Stylianos Bakas, Siegfried Ludwig, Konstantinos Barmpas, Mehdi Bahri, Yannis Panagakis, Nikolaos Laskaris, Dimitrios A. Adamos, Stefanos Zafeiriou
\\
\\
% Covariate shift
%Decoding of EEG signals suffers from substantial covariate shift between different subjects, sessions and datasets. This makes it difficult to find a common decoding model that can classify EEG signals with high accuracy and subject-independence. While deep learning models offer improved generalization given the increasing size of training datasets, it is further beneficial to address covariate shift explicitly.
% EEGInception architecture
Deep learning models based on the EEGInception architecture were used for both tasks \cite{santamaria2020eeg}. The architecture applies some modifications to the widely used EEGNet architecture \cite{lawhern2018eegnet}, including temporal filterbanks with different kernel sizes, as well as additional deeper processing layers. The convolutional layers are followed by a linear classifier. All models are trained with the Adam gradient descent optimizer \cite{kingma2014adam}, using default hyperparameters.

% Sleep task
As the sleep classification task consists of subject-independent classification on a single data set, a straight-forward approach of training a model on the combined source and calibration data is performed. Latent alignment between subjects is used as described in later paragraphs.

% MI dataset transfer
Beyond aligning different subjects, the motor imagery task requires the transfer of trained models from one or more source data sets to perform classification on two target data sets. To make data set transfer simple, only the Physionet MI source data set was used and the 30 common electrodes between the source and the two target data sets were selected. The models are trained on the source and calibration data simultaneously, using a shared feature extractor and separate linear classifier heads per data set \cite{huang2013cross, wei2021inter}.
% \begin{figure}[htbp]
 % Caption and label go in the first argument and the figure contents
 % go in the second argument
% \floatconts
%   {fig:image}
%   {\caption{Example Image}}
%   {\includegraphics[width=0.5\linewidth]{example-image}}
% \end{figure}

\begin{figure}[htbp]
\floatconts
    {fig:alignment}
    {\caption{Latent alignment is performed following feature extractor $\phi$ and the classification function $f$ is applied on the aligned features. a) Statistical alignment standardizes latent distributions of each subject. b) Deep Set alignment uses a trainable function $\Gamma$ to obtain a distribution embedding and update the features of each trial with a trainable function $\Lambda$. Source: \cite{bakas2022cogitat}. Solution proposed by team Cogitat.}}
    {\includegraphics[width=0.85\textwidth]{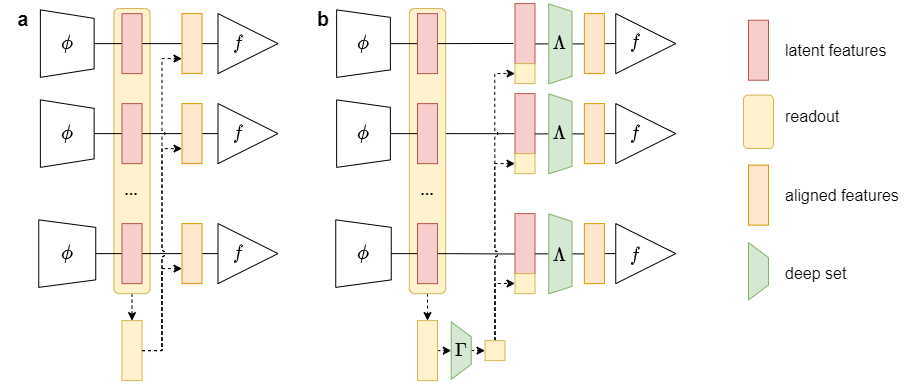}}
\end{figure}

% Latent alignment
The approach for improving subject-independence is to align the latent feature distributions of the deep learning models between different subjects and sessions \cite{bakas2022cogitat}. This is similar to the Euclidean Alignment method, which performs spatial whitening with the average covariance matrix per subject \cite{he2019transfer}, although the methods are not constrained to be applied on the model input. Performing distribution alignment in the classifier latent space is related to Riemannian manifold methods \cite{zanini2017transfer}, which have gained traction in recent years, but integration with deep learning models has not yet matured.

% Statistical alignment
For the first implementation of latent alignment, team Cogitat developed a novel statistical alignment method, which standardizes latent feature distributions per subject and can be seen as a subject-wise batch normalization \cite{li2021subject}, although a separate module per subject is not used (figure \ref{fig:alignment}a). This method is very computationally efficient, straightforward to implement and can be used in the place of the batch normalization layers in a deep learning model. During training, multiple trials need to be present for each subject in the batch, in order to estimate feature distributions. During inference, a statistical estimation obtained on unlabelled trials of the target subject can be used \cite{xu2021improving}. This worked well on the sleep classification task, where no subject-wise labelled calibration data is available.

% Deep sets
Taking the perspective of classification on a set of EEG trials from a given subject, the application of Deep Set architectures follows naturally as a second implementation of latent alignment \cite{zaheer2017deep}. Team Cogitat developed a deep learning layer that uses the statistical mean to obtain a distribution embedding of the given subject to update trial-wise features (figure \ref{fig:alignment}b). This worked particularly well on the MI task, which provides subject-wise calibration data.

\subsection{Second Place Solution: Multi-Source EEGNet with Domain and Label Adaptation}

Team \verb|wduong|: William C. Duong, Stephen M. Gordon and Vernon J. Lawhern \\

Team Wduong used a modified version of DeepSleep in Task 1 \cite{chambon2018deep} with some minor changes, including adjusting the convolution kernel size, adding batch normalization after each layer, and setting up a pre-training scheme. For the motor imagery decoding (Task 2), team Wduong used a combination of data alignment, multi-task EEGNet model, and maximum classifier discrepancy domain adaptation to train a robust model. The code is available at \url{https://github.com/mcd4874/NeurIPS_competition}.

\begin{figure}[htbp]
\floatconts
  {fig:EA}
  {\caption{Data sets alignment with Euclidean Alignment. Solution proposed by team Wduong.}}
  {%
    \subfigure[before Euclidean Alignment][c]{\label{fig:a}%
      \includegraphics[width=0.4\linewidth]{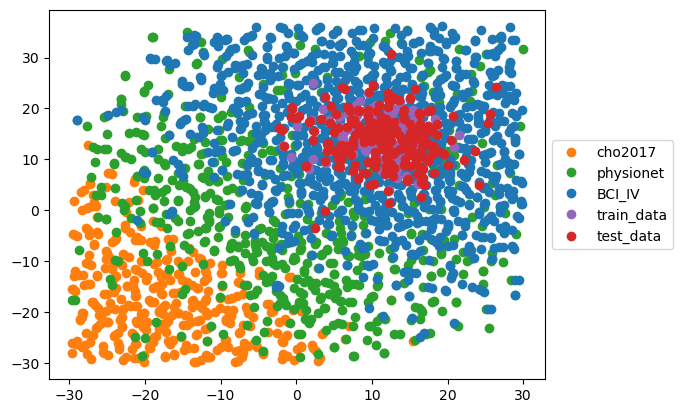}}%
    \qquad
    \subfigure[after Euclidean Alignment][c]{\label{fig:b}%
      \includegraphics[width=0.33\linewidth]{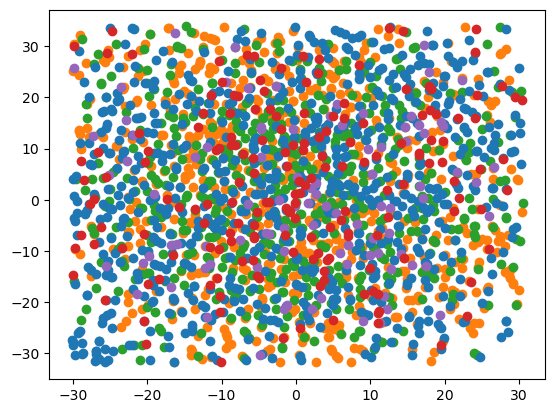}}
  }
%   \label{fig:EA}
\end{figure}

% \begin{figure}
%     \centering
%     \subfloat[
%         \centering before Euclidean Alignment
%     ]
%     {\includegraphics[scale=0.4]{BEETL/datasets_before_EA.png}\label{fig:a}}
%     \subfloat[
%         \centering after Euclidean Alignment
%     ]{\includegraphics[scale=0.4]{BEETL/datasets_after_EA.png}\label{fig:b}}
%     \caption{Data sets alignment with Euclidean Alignment. Solution proposed by team Wduong.}
%     \label{fig:EA}
% \end{figure}

Due to limited available labeled trials, the motor imagery decoding task needs transfer learning strategies to train a robust model with good performance. To create a generalized framework to handle multi-source data set transfer learning, all three provided data sets, included Cho2017, Physionet, and BNCI2014, are used in the training procedure. 17 common channels among three source data sets and two target data sets are utilised. To demonstrate the generalization of the approach, the tongue category in the BNCI2014 data sets is kept, which is not available in two target data sets. Two main problems exist among source data sets and target data sets. First, there is a category gap problem due to different label categories between source and target data sets. Label Alignment is applied to solve this problem \cite{he2020latransfer}. For example, the tongue motor imagery category can be projected to the rest category to align corresponding distributions; therefore, the training phase can use source data with different labels compared to target data. In addition to the label gap problem, other data sets with different experimental recording setups led to a domain gap between subjects from source data sets and target data sets. Euclidean Alignment is used to close the gap between subjects in source domains and target domain \cite{he2019transfer}. All trials of each subject are aligned such that the mean covariance matrices of all subjects are equal to the identity matrix after alignment. After alignment, data set distribution in both source domains and target domain are more similar as seen in figure \ref{fig:EA}.

Finally, a multi-task EEGNet combined with Maximum Classifier Discrepancy (MCD) \cite{mcd} is developed to solve the motor-imagery decoding problem. A multi-task learning (MTL) setup can learn a common EEGNet backbone to increase feature representation generalization among data sets, where MTL treats each data set from source domains and target domain as an individual task \cite{ruder2017overview}. Combining losses with a weighted sum can optimize the shared EEGNet backbone and all the classifiers jointly \cite{lawhern2018eegnet}. An adversarial training procedure in the MCD method is conducted between the target classifier and EEGNet backbone to both increase the discriminative ability of target classifiers and enhance the feature extraction of the backbone \cite{mcd}. The model architecture is in Figure \ref{fig:EEGNet_model}.

\begin{figure}[htbp]
\floatconts
    {fig:EEGNet_model}
    {\caption{Multi-task EEGNet. Solution proposed by team Wduong.}}
    {\includegraphics[width=0.78\textwidth]{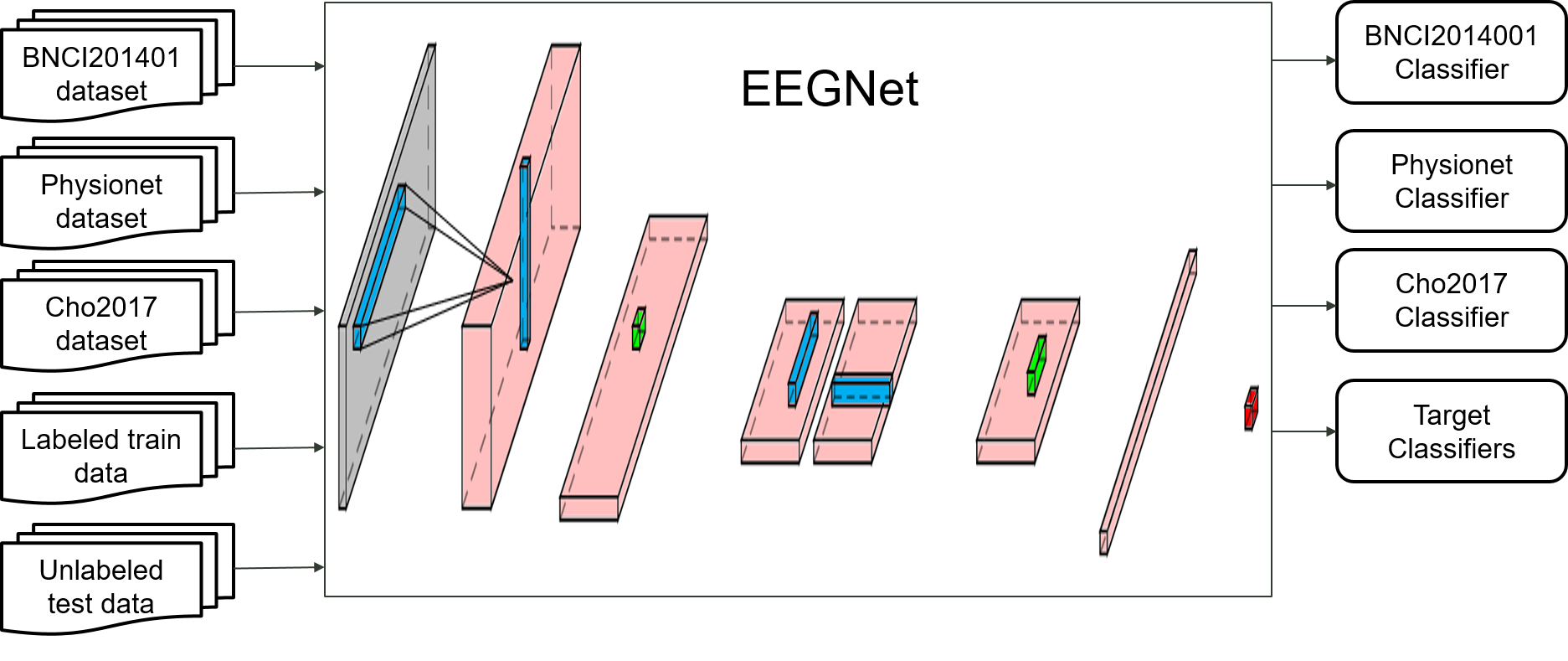}}
\end{figure}

% \begin{figure}[!ht]
%     \centering
%     \includegraphics[width=0.85\textwidth]{BEETL/wduong_beetl_figure.png}
%     \caption{Multi-task EEGNet }
%     \label{fig:EEGNet_model}
% \end{figure}

Each targeted subject is treated as a separate target domain for the training strategy due to the highly subject-specific nature of motor imagery EEG signals. Therefore, five model groups known as \(A_0, A_1, A_2, B_0\), and \(B_1 \) were created. Five-fold block-wise cross-validation is applied for each model group, such that 80\% of the target domain data is used to train a model and 20\% is used for validation to pick the best model. An ensemble learning strategy combines five-fold models via majority vote to learn the final prediction. During the training phase, the target data set samples a batch size of 16, while each source data set samples a batch size of 64 for each mini-batch to train the model. An Adam optimizer with the learning of 0.001 is used to train the model for 20 epochs.

\subsection{Third place solution: Classification of covariance matrices with SPDNet}

Team \verb|ms01|: Maciej Śliwowski, Vincent Rouanne, Piotr Tempczyk
\\
\\
Team ms01 performed different methods for task 1 and task 2 because of the difference in the available number of EEG channels. In the sleep decoding task (task 1), a DeepSleep model \cite{chambon2018deep} was trained and evaluated without major modifications. For the motor imagery decoding task (task 2), methods analyzing spatial covariance matrices of the signal was studied with a focus on Riemannian geometry approaches.

\begin{figure}[htbp]
\floatconts
  {fig:ms01}
  {\caption{Solution proposed by team ms01.}}
  {%
    \subfigure[Visualization of covariance matrices with t-SNE of small calibration data set for the second phase leaderboard data.][c]{\label{fig:ms01a}%
      \includegraphics[width=0.4\linewidth]{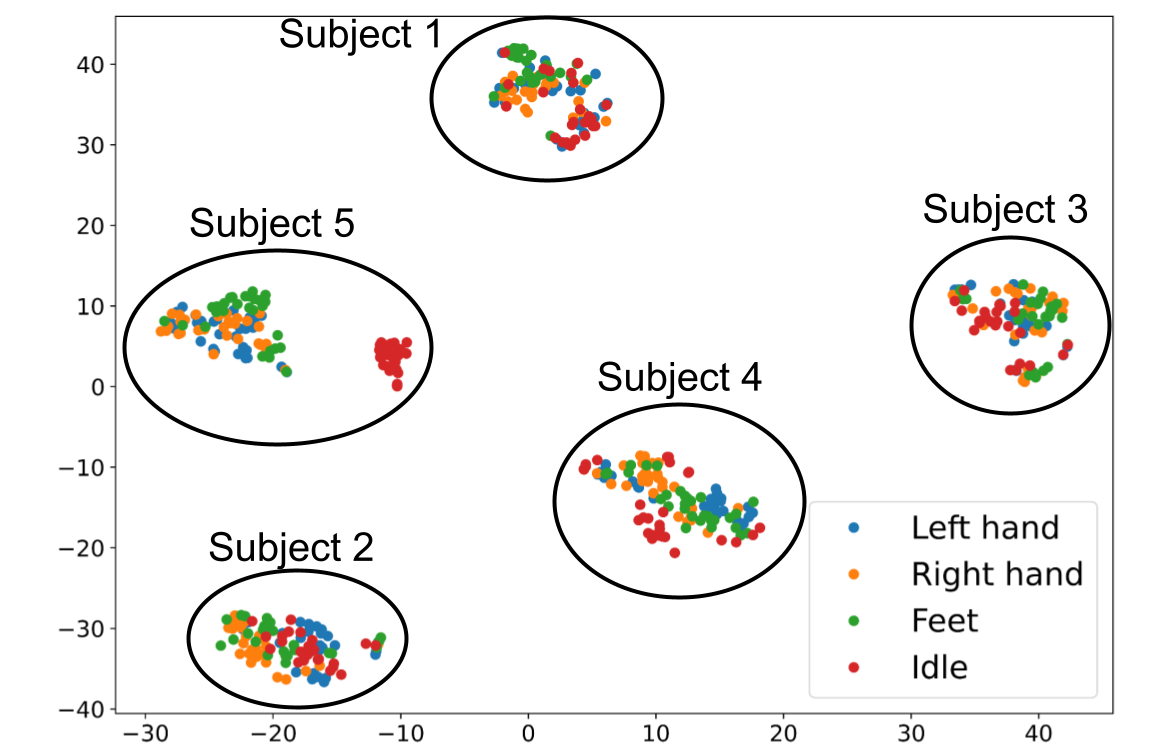}}%
    \qquad
    \subfigure[SPDNet visualization.][c]{\label{fig:ms01b}%
      \includegraphics[width=0.4\linewidth]{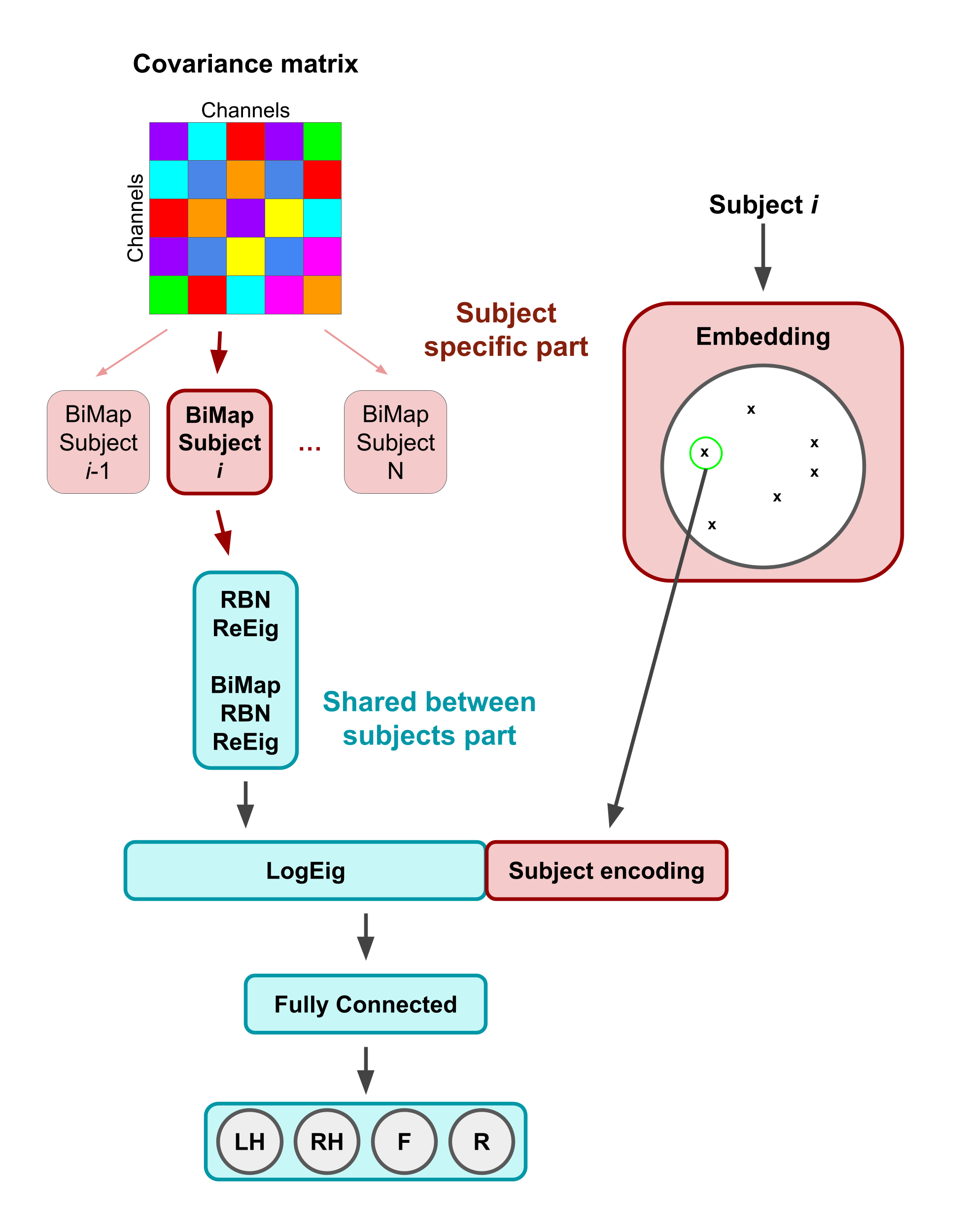}}
  }
\end{figure}

% \begin{figure}%
%     \centering
%     % \par\vspace{-5cm}\par
%     \raisebox{2cm}{\subfloat[\centering Visualization of covariance matrices with t-SNE of small calibration data set for the second phase leaderboard data.]{{\includegraphics[width=0.5\textwidth]{tsne_test_subjects.png}}}}%
%     \subfloat[\centering SPDNet visualization.]{{\includegraphics[width=0.45\textwidth]{diagram-spdnet.png}}}%
%     \caption{Solution proposed by team ms01.}%
%     \label{fig:ms01}%
% \end{figure}

% MDRM
The approach performs classification of the spatial covariance matrices for different motor imagery (MI) tasks with Minimum Distance to Riemannian Mean (MDRM) classifiers \cite{barachant_mdrm}. Covariance matrices lie on a manifold of positive semi-definite matrices in which the distance between data points can be computed using Riemannian geometry. The Riemannian mean of covariance matrices is computed separately for each motor imagery task. New observations are classified using the minimal distance between the observation's covariance matrix and the tasks mean covariance matrices. To visualize the data distribution and interpret the results obtained with the MDRM classifier, t-Stochastic Neighbor Embedding (t-SNE) \cite{vandermaaten_tsne} with Riemannian distance are used (Figure \ref{fig:ms01}a).

%SPDNet
Based on the MDRM results and the t-SNE visualization (Figure \ref{fig:ms01}a), the solution is proposed using nonlinear methods designed specifically for the analysis of symmetric positive definite matrices (SPD). SPD networks, similar to standard deep learning, are able to learn an efficient representation of the data that is not linearly separable as in our case. Covariance matrices are classified using SPDNet \cite{spdnet}, a type of Symmetric Positive Definite manifold networks, together with Riemannian batch normalization (RBN) \cite{riemannian_batch_norm} (code from \cite{riemannian_batch_norm}). All subjects from all sources and leaderboard labelled data sets are combined to form the training and validation data sets (30\% of the leaderboard labelled data set was used as validation). 

% SPDNet transfer learning
In the proposed SPDNet architecture for transfer learning (Figure \ref{fig:ms01}b), data for each subject is processed by an individual feature extractor consisting of one bilinear transformation (BiMap layer) \cite{spdnet}.
As further steps in the model are shared between subjects, the first BiMap layer is able to unify the representations and make it subject invariant. The shared part of the network consists of RBN and rectified eigenvalues activation (ReEig) \cite{spdnet} followed by another block of BiMap, RBN, and ReEig. Extracted features are transformed from Riemannian space into Euclidean space (LogEig layer). Two fully connected layers (with a dropout layer in-between) are added to predict four MI classes. To catch between-subject variance, an embedding/encoding of patients of size fifteen is created, which is concatenated to the first fully connected euclidean layer. The embedding is trained together with the whole network with backpropagation. A hypothesis is that the subject encoding could ease features standardization across subjects.

\section{Discussion}
\label{sec:discussion}
% Introduction
In this section, we will discuss key observations from the BEETL competition. This section includes the reasoning for choices of algorithms, the commonalities and differences among solutions, limitations of the winning methods and task design, and future directions.

% model architecture
Architecture design is a challenge for cross-dataset EEG transfer learning. In light of the relatively large data sets provided, team Cogitat chose the EEGInception architecture for both tasks, which has enough capacity to learn discriminative features across subjects and data sets. Both team Wduong and ms01 used a modified DeepSleep model for task 1. In task 2, team Wduong used a multi-task EEGNet setup. Team ms01 used the SPDNet architecture that analyzes covariance matrices with nonlinear transformations. Despite the diversity of algorithms, all teams used deep learning approaches. The choice of using neural networks might be explained by the need for increased model capacity when attempting to find classifiers that can generalize across many subjects and data sets. Further, deep learning architectures allow for end-to-end training of various processing stages, including representation learning without hand-crafted features and distribution alignment. This provides flexibility in the design of architectures to test out different transfer learning setups. 

%Split branches for each dataset
Another commonality is that all employed a combination of shared network backbones and individualized processing layers for different data sets or subjects (either in shallow layers or classification layers). Team Cogitat and Wduong used a shared feature extractor with different classifiers for different data sets. Team ms01 had unique layers for each subject at shallow layers of their architecture. The splitting approach allows model backbones to be trained on large amounts of data while allowing for adaptation to the existing differences. 

% subject alignment
EEG decoding suffers from substantial covariate shifts among different subjects and data sets. All three teams used alignment approaches to reduce differences among data sets, subjects and sessions. Team Cogitat performed statistical alignment of latent features at various stages of the model. Team Wduong performed Euclidean alignment on the EEG trials based on their covariances and an additional label alignment approach. Team ms01 analyzed covariance matrices with individual per-subject layers to unify the representations. Although the exact implementations vary, the winning solutions indicate the utility of alignment methods to perform domain adaptation in EEG transfer learning methods.

% Choosing source data sets
How to select and process source data sets is a key question for EEG transfer learning towards the use of big data. In task 2, team Wduong and team ms01 used all three source data sets with 17 common channels to utilise more subjects and tasks. Team Cogitat used one source data set (PhysionetMI) with more electrodes. Here we highlight a trade-off between keeping more common electrode channels or increasing the number of subjects by adding more data sets. % channel selection
One limitation of the winning solutions is that they either abandoned some channels or some subjects. Discriminative information could be lost by this selection approach. Therefore, aiming to apply some form of upsampling approach or building a better architecture to use all channels of data sets would be possible future directions.

We observe that the gaps of decoding accuracies between participants are relatively small in the sleep task. Most teams did not utilise complex domain adaptation strategies. This may be caused by various reasons. Firstly, sleep EEG could be relatively similar across subjects by nature compared to motor imagery. Secondly, the sleep task is designed on only a single data set with two channels, which simplified the scenario for medical diagnostics. Another human factor could be that sleep data is manually labelled by human experts which unified the data in terms of classes. More studies on utilising multiple sleep data sets with more channels and setups are needed in the literature to conclude a precise explanation. 

There are various directions that require more studies. In the competition, backbone architectures are based on similar ideas of temporal and spatial filters \cite{lawhern2018eegnet, santamaria2020eeg}. More architectures in the EEG field should be explored for multi-dataset transfer learning, including conventional transfer learning algorithms \cite{8xx}. Winning solutions used various alignment approaches, whereas a systematical comparison of these alignment approaches in the same experiment setup is required. In addition, the winning methods either leave out some channels or source tasks in their approaches. Therefore, algorithms that could use the full potential of sources should be developed in the future. Furthermore, in light of the fact that all winning solutions used deep learning, the interpretability \cite{goebel2018explainable, samek2019explainable} of features used for deep transfer learning needs to be further investigated.

To conclude, the BEETL competition brought attention to the challenge of using large-scale biosignal transfer learning on a large amount of subjects and data sets. It provides a common platform to evaluate transfer learning algorithms and some valuable examples in EEG transfer learning with insights. We hope the BEETL competition could open up the opportunity to utilise heterogenous open-source data sets and move the EEG field forward.

% The bibliography is displayed using \verb|\bibliography|.
\acks{
Our competition is officially affiliated with the BCI Society. We thank Facebook Reality Labs for sponsoring 5000\$ in competition prizes. Many thanks to NeurIPS2021 Organisers and all researchers who devote themselves into this competition. We thank all organisers and teams in BEETL. Aldo Faisal acknowledges a UKRI Turing AI Fellowship Grant (EP/V025449/1). Alexandre Gramfort acknowledges the support of ANR BrAIN AI chair (ANR-20-CHIA-0016). Maciej Śliwowski was supported by the CEA NUMERICS program, which has received funding from European Union’s Horizon 2020 research and innovation program under the Marie Sklodowska-Curie grant agreement No 800945.}

% \input{appendix.tex}
% \bibliography{reference}

\end{document}